# Psychological Mechanisms of Generative AI Discontinuance Intention among Chinese K–12 Teachers

**First Author and Corresponding Author**
Yiran Du
University of Cambridge, Cambridge, UK
yd392@cam.ac.uk

**Second Author**
Qian Chen
College of Arts and Social Sciences, Australian National University, Canberra, Australia
qianch2021@163.com

**Third Author**
Huimin He
Xi'an Jiaotong-Liverpool University, Suzhou, China
Huimin.he@xjtlu.edu.cn

**Abstract**
This study examines the psychological mechanisms underlying Chinese K–12 teachers' discontinuance intention toward generative AI. Drawing on the Cognition–Affect–Conation framework, the study investigates how cognitive evaluations of generative AI shape affective responses and subsequently influence behavioural intention. Survey data from 256 Chinese K–12 teachers were analysed using structural equation modelling and fuzzy-set qualitative comparative analysis. The results showed that privacy concern, algorithmic opacity, and information hallucination increased AI anxiety, which in turn strengthened discontinuance intention. Conversely, perceived intelligence, perceived personalisation, and perceived interactivity enhanced satisfaction, which reduced discontinuance intention. The configurational analysis further identified multiple pathways leading to high discontinuance intention, highlighting the combined roles of technological risks, AI anxiety, weak affordance perceptions, and low satisfaction. These findings extend research on post-adoption generative AI use in education and suggest that sustainable integration requires both reducing technological uncertainty and enhancing teachers' positive user experiences.

**Keywords:** generative AI; discontinuance intention; Cognition–Affect–Conation framework; AI anxiety; teacher technology use

## 1. Introduction

Generative AI, has rapidly entered educational practice, offering new opportunities for teachers to generate instructional materials, design assessments, provide feedback, and support lesson preparation (Qian, 2025; Tan et al., 2025). In K–12 education, generative AI is increasingly viewed as a tool that may reduce teachers' workload, improve instructional efficiency, and support differentiated teaching (Alfarwan, 2025; N. Wang et al., 2024). However, its classroom integration also raises concerns about inaccurate outputs, data privacy, algorithmic bias, and ethical risks, which may affect teachers' willingness to use such tools continuously (Kasneci et al., 2023; Mao et al., 2024; Shata, 2025).

Although prior studies have examined teachers' adoption, readiness, and continuance intention regarding artificial intelligence and generative AI in education (Ayanwale et al., 2025; Chou et al., 2025; Zheng et al., 2025), less attention has been paid to discontinuance intention, particularly after teachers have already experimented with generative AI in teaching. Existing discontinuance research suggests that users may abandon a technology when perceived risks, emotional discomfort, or unsatisfactory experiences outweigh perceived benefits (Lin et al., 2020; Zhou & Wang, 2025; Zhou & Zhang, 2025). In the context of Chinese K–12 teaching, this issue is especially important because teachers must balance instructional innovation with professional responsibility, student data protection, and the accuracy of teaching content.

To address this gap, the present study examines the psychological mechanisms underlying Chinese K–12 teachers' generative AI discontinuance intention. Drawing on the Cognition–Affect–Conation framework, this study investigates how teachers' cognitive evaluations of generative AI, including privacy concern, algorithmic opacity, information hallucination, perceived intelligence, perceived personalisation, and perceived interactivity, shape affective responses, namely AI anxiety and satisfaction, which in turn influence discontinuance intention (Qaisar et al., 2024; Zeng et al., 2023). By combining structural equation modelling (SEM) and fuzzy-set qualitative comparative analysis (fsQCA), this study aims to explain both the net effects and configurational pathways through which teachers develop intentions to reduce or stop using generative AI in teaching.

## 2. Literature Review

### 2.1 Generative AI in Teaching

Generative AI refers to artificial intelligence systems capable of producing new content, such as text, images, or code, based on patterns learned from large datasets (Y. Wang et al., 2025). With the emergence of large language models and multimodal generative systems, generative AI has begun to influence educational practices worldwide (Nzenwata et al., 2024). In China, the rapid development of domestic and international generative AI tools has stimulated growing interest in their potential applications within K–12 teaching contexts (Lee et al., 2026).

Recent literature suggests that generative AI can support teachers in multiple instructional tasks, including lesson planning, material development, assessment design, and feedback generation (Qian, 2025). By automating routine cognitive tasks, these tools may reduce teachers' workload and enhance instructional efficiency (N. Wang et al., 2024). In the Chinese K–12 system, where teachers often face heavy teaching and administrative responsibilities, GenAI has been viewed as a potential means of improving teaching productivity and supporting differentiated instruction (Alfarwan, 2025).

Despite these potential benefits, scholars also highlight several challenges associated with integrating generative AI into teaching (B. Li et al., 2024). Concerns related to content accuracy, algorithmic bias, ethical considerations, and data security have generated uncertainty among educators (Shata, 2025). Within the Chinese educational context, where instructional quality and examination performance are highly emphasised, such concerns may shape teachers' perceptions and emotional responses toward the continued use of generative AI tools (Mao et al., 2024).

### 2.2 Discontinuance Intention

Discontinuance intention generally refers to an individual's conscious intention to stop using a technology after a period of prior use (Lin et al., 2020). In information systems research, it represents a post-adoption behavioural tendency in which users reassess the value, risks, or outcomes of a system and decide whether to abandon it (Zhou & Wang, 2025). This concept contrasts with adoption or continuance intention, which focus on the initiation or sustained use of technology (Zhou & Li, 2025).

In the context of generative AI in education, teachers' discontinuance intention can be defined as their deliberate intention to reduce or completely stop using generative AI tools in their teaching activities after initial experimentation or adoption (Ayanwale et al., 2025). This intention may arise when teachers perceive that the technology does not meet their instructional needs, introduces pedagogical risks, or generates negative experiences during classroom use (Kasneci et al., 2023).

Prior research indicates that discontinuance intention is influenced by a combination of cognitive evaluations and affective responses (Zhou & Wang, 2025). Teachers may develop discontinuance intentions when perceived risks outweigh perceived benefits, when the technology conflicts with established teaching practices, or when emotional reactions such as frustration, anxiety, or distrust emerge (Zhou & Zhang, 2025). Understanding these factors is essential for explaining whether generative AI can be sustainably integrated into Chinese K–12 educational practice.

## 3. Theoretical Framework and Hypothesis Development

### 3.1 Cognition–Affect–Conation Framework

The Cognition–Affect–Conation (CAC) framework explains behavioural intention as a sequential psychological process in which individuals' cognitive evaluations influence emotional reactions, which subsequently shape behavioural responses (Zeng et al., 2023). The framework has been widely applied in behavioural science, psychology, and information systems research to explain how users develop intentions toward adopting, continuing, or discontinuing technology use (Qaisar et al., 2024). Applying this framework is appropriate in the context of generative AI in education because teachers' behavioural decisions are unlikely to arise solely from rational evaluation; rather, they emerge from the interaction between cognitive perceptions of the technology and affective responses generated during use (Y. Du, Tang, et al., 2026). When teachers interact with generative AI tools in their instructional practice, they first evaluate the characteristics of the technology, such as its potential risks and capabilities. These cognitive evaluations then evoke emotional responses, including feelings of anxiety or satisfaction, which subsequently shape teachers' behavioural intentions regarding whether to continue or discontinue using the technology.

In this study, the CAC framework provides a theoretically grounded structure for explaining Chinese K–12 teachers' discontinuance intention toward generative AI. As illustrated in the conceptual model (see Figure 1), teachers' cognitive evaluations of generative AI are conceptualised from two complementary perspectives: enablers and inhibitors. This dual perspective reflects the notion that users' post-adoption decisions are shaped by the simultaneous consideration of both perceived risks and perceived benefits of a technology (Zhou & Wang, 2025). Enablers capture negative technological perceptions that may facilitate discontinuance by generating adverse emotional responses. In this study, these include privacy concern, algorithmic opacity, and information hallucination, which are expected to increase teachers' AI anxiety. In contrast, inhibitors represent positive perceptions of generative AI capabilities that may discourage discontinuance by fostering favourable emotional experiences. These include perceived intelligence, perceived personalisation, and perceived interactivity, which are expected to enhance teachers' satisfaction with the technology. At the affective level, AI anxiety and satisfaction represent teachers' emotional responses toward generative AI. These affective states subsequently influence the conative stage, where teachers develop discontinuance intention, with anxiety expected to increase discontinuance intention and satisfaction expected to reduce it.

**Figure 1. The Conceptual Model**

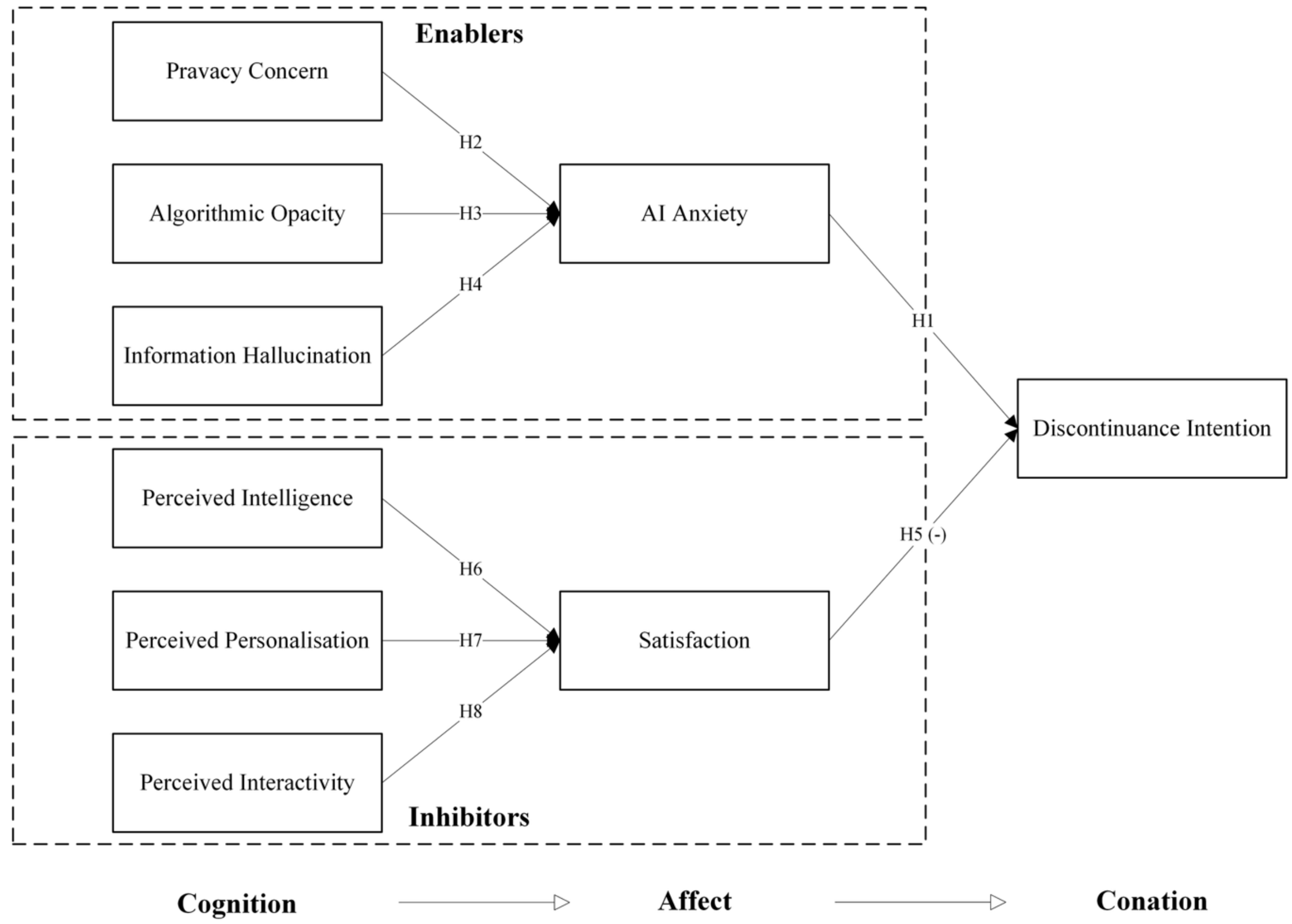

### 3.2 AI Anxiety, Privacy Concern, Algorithmic Opacity, and Information Hallucination

AI anxiety refers to users' feelings of apprehension, uncertainty, or discomfort when interacting with artificial intelligence technologies (Johnson & Verdicchio, 2017). Prior research in human–computer interaction and automation has shown that individuals may experience anxiety when technologies appear autonomous, unpredictable, or difficult to control (Kaya et al., 2024). Empirical studies further suggest that technology-related anxiety can significantly influence users' behavioural responses, including avoidance and discontinuance of digital systems (Wu & Li, 2025). In educational contexts, teachers who experience higher levels of AI anxiety may hesitate to rely on generative AI tools in instructional activities, which may increase their intention to discontinue using such technologies (J. J. H. Kim et al., 2025).

Privacy concern represents an important cognitive antecedent of AI anxiety (J. Li & Huang, 2020). In information systems research, privacy concern is commonly defined as users' apprehension regarding the collection, storage, and potential misuse of personal data when interacting with digital systems (Herriger et al., 2025). Empirical evidence indicates that heightened privacy concerns can increase perceived technological risks and generate negative emotional responses toward emerging technologies (Carmody et al., 2021; Hu & Min, 2023). For teachers using generative AI, concerns may arise regarding whether instructional materials, classroom interactions, or student-related information entered into AI systems could be stored or utilised without sufficient control, thereby increasing anxiety during system use (Zheng et al., 2025).

Algorithmic opacity also contributes to negative psychological responses toward AI systems (Guo et al., 2025). Algorithmic opacity refers to the lack of transparency in how algorithms process data and generate outputs (Yang et al., 2024). When users cannot clearly understand how an AI system produces responses, they may perceive the technology as unpredictable or uncontrollable (Vaassen, 2022). Prior studies in algorithmic decision-making and explainable AI indicate that opaque systems often reduce user trust and increase uncertainty (Eslami et al., 2019). For teachers, limited transparency regarding how generative AI generates teaching content may heighten anxiety about relying on such systems for instructional purposes (B. Li et al., 2025).

Information hallucination represents another critical issue associated with generative AI systems. Hallucination refers to the generation of plausible but factually incorrect or fabricated information by AI models (Sun et al., 2024). Recent empirical research on large language models has documented that such systems may produce inaccurate or misleading outputs despite appearing confident (Huang et al., 2025). In educational settings, the risk of using incorrect AI-generated content may create concerns about instructional accuracy and professional responsibility among teachers (Y. Du, Tang, et al., 2026). These concerns may further intensify AI anxiety when teachers interact with generative AI tools (Zhou & Zhang, 2025). Accordingly, the following hypotheses are proposed:

H1: AI anxiety is positively associated discontinuance intention to use generative AI.
H2: Privacy concern is positively associated with AI anxiety.
H3: Algorithmic opacity is positively associated with AI anxiety.
H4: Information hallucination is positively associated with AI anxiety.

### 3.3 Satisfaction, Perceived Intelligence, Perceived Personalisation, and Perceived Interactivity

Satisfaction refers to users' overall affective evaluation of their experience with a technology after interacting with it (Hoffman et al., 2023). In information systems research, satisfaction is considered a key predictor of post-adoption behaviour. Empirical studies indicate that satisfied users are more likely to continue using a technology, whereas dissatisfaction often leads to discontinuance (Xie et al., 2024). In the context of generative AI in teaching, teachers who perceive the technology as useful and effective in supporting instructional tasks are more likely to develop positive experiences, thereby reducing their intention to discontinue its use (Tan et al., 2025).

Perceived intelligence refers to the extent to which users believe that an AI system demonstrates competent and knowledgeable responses (Tusseyeva et al., 2024). Generative AI systems can produce explanations, instructional materials, and feedback that resemble human-like reasoning (Chou et al., 2025). When teachers perceive these outputs as accurate, relevant, and contextually appropriate, they are more likely to evaluate the system positively (C. Zhang et al., 2023). For example, prior research suggests that higher perceived intelligence enhances users' confidence in AI systems and contributes to greater satisfaction (Zhou & Zhang, 2024).

Perceived personalisation reflects the degree to which users believe that a system can tailor responses to their individual needs or contexts (Niu et al., 2026). In educational environments, personalised technologies can provide customised explanations, examples, or instructional suggestions (Law, 2024). When generative AI systems adapt outputs according to teachers' subject areas or teaching goals, teachers may perceive the system as more supportive and responsive, thereby increasing their satisfaction with the technology (Al Darayseh, 2023).

Perceived interactivity refers to the extent to which users perceive a system as responsive and capable of facilitating dynamic interaction (F. Wang et al., 2025). Generative AI systems allow users to iteratively refine prompts and receive immediate feedback through conversational exchanges (C. Wang et al., 2024). Studies on interactive technologies suggest that higher perceived interactivity enhances user engagement and positive experience (J. Kim et al., 2022). In teaching contexts, responsive interaction with generative AI tools may therefore contribute to teachers' satisfaction with the system (Xiang & Chae, 2022). Accordingly, the following hypotheses are proposed:

H5: Satisfaction is negatively associated with discontinuance intention to use generative AI.
H6: Perceived intelligence is positively associated with satisfaction.
H7: Perceived personalisation is positively associated with satisfaction.
H8: Perceived interactivity is positively associated with satisfaction.

## 4. Methods

### 4.1 Research Design

This study adopted a quantitative, cross-sectional survey design to examine the psychological mechanisms underlying Chinese K–12 teachers' discontinuance intention toward generative AI. Guided by the Cognition–Affect–Conation framework, the study tested a structural model in which teachers' cognitive evaluations of generative AI, including privacy concern, algorithmic opacity, information hallucination, perceived intelligence, perceived personalisation, and perceived interactivity, shaped affective responses, namely AI anxiety and satisfaction, which in turn predicted discontinuance intention. Survey data were collected from teachers with prior experience using generative AI in teaching, and the proposed relationships were analysed using structural equation modelling and fuzzy-set qualitative comparative analysis to capture both net effects and configurational mechanisms.

### 4.2 Participants

The quantitative phase involved 256 Chinese K–12 teachers who had experience using generative AI tools in their teaching practice, after excluding 34 careless responses (Ward & Meade, 2023). Participants were recruited through online teacher communities and professional networks. Table 1 summarises the demographic characteristics of the respondents. Among the participants, 23.4% were under 25 years old (*n* = 60), 43.0% were between 25 and 34 years old (*n* = 110), 21.9% were between 35 and 44 years old (*n* = 56), and 11.7% were aged 45 or above (*n* = 30). The sample included 110 male teachers (43.0%) and 146 female teachers (57.0%). In terms of teaching subjects, 120 participants (46.9%) taught STEM-related subjects, while 136 (53.1%) taught non-STEM subjects. Regarding teaching experience, 27.3% had less than three years of experience (*n* = 70), 25.0% had three to five years (*n* = 64), 26.6% had six to ten years (*n* = 68), and 21.1% had more than ten years of experience (*n* = 54). Participants were drawn from different school levels, including primary schools (35.2%, *n* = 90), junior secondary schools (34.4%, *n* = 88), and senior secondary schools (30.5%, *n* = 78). Overall, the survey sample reflected teachers with varied demographic and professional backgrounds.

**Table 1. Participant Characteristics (*N* = 256)**

| Variable | Category | Frequency (*n*) | Percentage (%) |
|---|---|---|---|
| Age | Under 25 | 60 | 23.4 |
| | 25–34 | 110 | 43.0 |
| | 35–44 | 56 | 21.9 |
| | 45 or above | 30 | 11.7 |
| Gender | Male | 110 | 43.0 |
| | Female | 146 | 57.0 |
| Teaching subject | STEM | 120 | 46.9 |
| | Non-STEM | 136 | 53.1 |
| Teaching experience | Less than 3 years | 70 | 27.3 |
| | 3–5 years | 64 | 25.0 |
| | 6–10 years | 68 | 26.6 |
| | More than 10 years | 54 | 21.1 |
| Teaching level | Primary | 90 | 35.2 |
| | Junior secondary | 88 | 34.4 |
| | Senior secondary | 78 | 30.5 |

### 4.3 Measurement

All constructs, privacy concern (Zhou & Zhang, 2025), algorithmic opacity (Guo et al., 2025), information hallucination (Zhou & Li, 2025), perceived intelligence (Zhou & Wang, 2025), perceived personalisation (Zhou & Wang, 2025), perceived interactivity (Zhou & Zhang, 2024), AI anxiety (Kaya et al., 2024), satisfaction (Ashfaq et al., 2020), discontinuance intention (Lin et al., 2020), were measured using established multi-item scales adapted from prior literature and tailored to the current context (Y. Du, 2024). All items were rated on a five-point Likert scale ranging from 1 (strongly disagree) to 5 (strongly agree). The items were originally developed in English and translated into Chinese following a back-translation procedure to ensure semantic equivalence (Brislin, 1970). Prior to the main survey, the instrument was piloted with a small group of teachers (*N* = 30) to assess item clarity and contextual appropriateness, and minor wording adjustments were made accordingly. Because the data were obtained entirely from a self-report questionnaire, common method variance (CMV) was examined (Podsakoff et al., 2024). Harman's single-factor test indicated that the first unrotated factor explained 23.6% of the total variance, suggesting that CMV was unlikely to pose a substantial threat. The full list of measurement items is presented in Table 2.

**Table 2. Constructs and Measurement Items**

| Construct | Code | Measurement Item (English) | Measurement Item (Chinese) |
|---|---|---|---|
| Privacy Concern (PC) | PC1 | I am concerned about the privacy of student information when using generative AI tools. | 在使用生成式人工智能工具时，我担心学生信息的隐私安全。 |
| | PC2 | I worry that using generative AI in teaching may expose sensitive classroom data. | 我担心在教学中使用生成式人工智能可能会暴露课堂中的敏感数据。 |
| | PC3 | I feel uneasy about sharing teaching materials with generative AI systems. | 将教学材料输入生成式人工智能系统让我感到不安。 |
| Algorithmic Opacity (AO) | AO1 | I do not understand how generative AI produces its answers. | 我不理解生成式人工智能是如何生成其回答的。 |
| | AO2 | The reasoning process of generative AI systems is unclear to me. | 生成式人工智能系统的推理过程对我来说是不清晰的。 |
| | AO3 | It is difficult for me to explain how generative AI generates teaching content. | 我很难解释生成式人工智能是如何生成教学内容的。 |
| Information Hallucination (IH) | IH1 | Generative AI sometimes produces inaccurate teaching information. | 生成式人工智能有时会生成不准确的教学信息。 |
| | IH2 | The content generated by generative AI is sometimes unreliable. | 生成式人工智能生成的内容有时不够可靠。 |
| | IH3 | Generative AI may generate explanations that are not factually correct. | 生成式人工智能有时会生成与事实不符的解释。 |
| Perceived Intelligence (PI) | PI1 | Generative AI demonstrates strong knowledge in teaching-related topics. | 生成式人工智能在教学相关问题上表现出较强的知识能力。 |
| | PI2 | Generative AI can understand my teaching questions well. | 生成式人工智能能够很好地理解我的教学问题。 |
| | PI3 | Generative AI provides competent responses to instructional questions. | 生成式人工智能能够对教学问题提供有效的回答。 |
| Perceived Personalisation (PP) | PP1 | Generative AI can tailor instructional content to my teaching needs. | 生成式人工智能能够根据我的教学需求调整教学内容。 |
| | PP2 | Generative AI can generate materials suitable for my classroom context. | 生成式人工智能能够生成适合我课堂情境的教学材料。 |
| | PP3 | Generative AI can adapt explanations for different learning situations. | 生成式人工智能能够针对不同学习情境调整解释方式。 |
| | PIN1 | Generative AI responds quickly to my teaching questions. | 生成式人工智能能够快速回应我的教学问题。 |
| Perceived Interactivity (PIN) | PIN2 | My interaction with generative AI is responsive. | 我与生成式人工智能的互动具有良好的响应性。 |
| | PIN3 | Generative AI provides interactive feedback during teaching preparation. | 在备课过程中，生成式人工智能能够提供互动性的反馈。 |
| AI Anxiety (AA) | AA1 | I feel uneasy when using generative AI for teaching tasks. | 在教学任务中使用生成式人工智能时，我感到不安。 |
| | AA2 | Using generative AI in teaching makes me feel nervous. | 在教学中使用生成式人工智能让我感到紧张。 |
| | AA3 | I feel worried when relying on generative AI in my teaching work. | 在教学工作中依赖生成式人工智能让我感到担忧。 |
| Satisfaction (SAT) | SAT1 | I am satisfied with my experience using generative AI in teaching. | 我对在教学中使用生成式人工智能的体验感到满意。 |
| | SAT2 | Using generative AI for teaching is satisfying. | 在教学中使用生成式人工智能让我感到满意。 |
| | SAT3 | Overall, I am pleased with generative AI tools for teaching tasks. | 总体而言，我对生成式人工智能在教学中的应用感到满意。 |
| Discontinuance Intention (DI) | DI1 | I intend to reduce my use of generative AI in teaching. | 我打算减少在教学中使用生成式人工智能。 |
| | DI2 | I plan to stop using generative AI for teaching tasks. | 我计划停止在教学任务中使用生成式人工智能。 |
| | DI3 | I may discontinue using generative AI in my teaching activities. | 我可能会停止在教学活动中使用生成式人工智能。 |

### 4.4 Data Analysis

The data were analysed using a two-stage approach. First, structural equation modelling (SEM) (Kline, 2023) was conducted to examine the measurement model and test the hypothesised relationships among the constructs. Descriptive statistics, skewness, and kurtosis were assessed to evaluate data distribution, while model fit was examined using $\chi^2$/df, CFI, TLI, RMSEA, and SRMR. Reliability and validity were evaluated through standardised factor loadings, Cronbach's alpha, composite reliability, average variance extracted, inter-construct correlations, and HTMT ratios. Second, fuzzy-set qualitative comparative analysis (fsQCA) (Pappas & Woodside, 2021) was employed to identify configurations of cognitive and affective conditions associated with high discontinuance intention. The variables were calibrated using the 95th, 50th, and 5th percentiles as anchors for full membership, crossover, and full non-membership, respectively, followed by necessity and sufficiency analyses.

## 5. Results

### 5.1 Structural Equation Modelling (SEM) Results

The SEM results indicated that both the measurement and structural models met the recommended threshold criteria. As shown in Table 3, the descriptive statistics showed no serious departure from normality, as skewness and kurtosis values were within the conventional ±2 range. The model fit indices were also acceptable. Specifically, $\chi^2$/df was below the threshold of 3.00, CFI and TLI exceeded the recommended value of .90, and RMSEA and SRMR were below the recommended value of .08. The measurement model showed good fit, $\chi^2$/df = 1.86, CFI = .96, TLI = .95, RMSEA = .06, and SRMR = .05, while the structural model also showed acceptable fit, $\chi^2$/df = 1.94, CFI = .95, TLI = .94, RMSEA = .06, and SRMR = .05 (Table 4). Reliability and convergent validity were supported, as all standardised factor loadings exceeded .70, Cronbach's $\alpha$ and composite reliability values were above .70, and AVE values were above .50 (Table 5). Discriminant validity was also established because the square roots of AVE were greater than the corresponding inter-construct correlations, and all HTMT ratios were below the conservative threshold of .85 (Tables 6 and 7).

**Table 3. Descriptive Statistics of the Constructs**

| Construct | *M* | *SD* | Skewness | Kurtosis |
|---|---|---|---|---|
| Privacy Concern | 3.62 | 0.84 | −0.41 | −0.36 |
| Algorithmic Opacity | 3.48 | 0.79 | −0.28 | −0.42 |
| Information Hallucination | 3.71 | 0.76 | −0.52 | −0.18 |
| Perceived Intelligence | 3.89 | 0.68 | −0.64 | 0.21 |
| Perceived Personalisation | 3.74 | 0.72 | −0.49 | −0.07 |
| Perceived Interactivity | 3.81 | 0.70 | −0.55 | 0.12 |
| AI Anxiety | 3.34 | 0.86 | −0.19 | −0.58 |
| Satisfaction | 3.76 | 0.73 | −0.47 | −0.11 |
| Discontinuance Intention | 3.08 | 0.91 | 0.16 | −0.61 |

**Table 4. Model Fit Indices**

| Fit index | Threshold | Measurement model | Structural model |
|---|---|---|---|
| $\chi^2/df$ | < 3.00 | 1.86 | 1.94 |
| CFI | > 0.90 | 0.96 | 0.95 |
| TLI | > 0.90 | 0.95 | 0.94 |
| RMSEA | < 0.08 | 0.06 | 0.06 |
| SRMR | < 0.08 | 0.05 | 0.05 |

**Table 5. Reliability and Convergent Validity**

| Construct | Item | Loading | Cronbach's α | CR | AVE |
|---|---|---|---|---|---|
| Privacy Concern | PC1 | 0.82 | 0.86 | 0.86 | 0.68 |
| | PC2 | 0.85 | | | |
| | PC3 | 0.80 | | | |
| Algorithmic Opacity | AO1 | 0.79 | 0.84 | 0.84 | 0.64 |
| | AO2 | 0.83 | | | |
| | AO3 | 0.78 | | | |
| Information Hallucination | IH1 | 0.84 | 0.88 | 0.88 | 0.71 |
| | IH2 | 0.87 | | | |
| | IH3 | 0.82 | | | |
| Perceived Intelligence | PI1 | 0.85 | 0.89 | 0.89 | 0.73 |
| | PI2 | 0.88 | | | |
| | PI3 | 0.83 | | | |
| Perceived Personalisation | PP1 | 0.82 | 0.87 | 0.87 | 0.69 |
| | PP2 | 0.86 | | | |
| | PP3 | 0.81 | | | |
| Perceived Interactivity | PIN1 | 0.80 | 0.85 | 0.85 | 0.66 |
| | PIN2 | 0.84 | | | |
| | PIN3 | 0.79 | | | |
| AI Anxiety | AA1 | 0.86 | 0.90 | 0.90 | 0.75 |
| | AA2 | 0.88 | | | |
| | AA3 | 0.85 | | | |
| Satisfaction | SAT1 | 0.87 | 0.91 | 0.91 | 0.77 |
| | SAT2 | 0.90 | | | |
| | SAT3 | 0.86 | | | |
| Discontinuance Intention | DI1 | 0.84 | 0.88 | 0.88 | 0.70 |
| | DI2 | 0.87 | | | |
| | DI3 | 0.80 | | | |

**Table 6. Reliability and Convergent Validity**

| Construct | PC | AO | IH | PI | PP | PIN | AA | SAT | DI |
|---|---|---|---|---|---|---|---|---|---|
| PC | **0.82** | | | | | | | | |
| AO | 0.51 | **0.80** | | | | | | | |
| IH | 0.48 | 0.54 | **0.84** | | | | | | |
| PI | −0.31 | −0.28 | −0.34 | **0.85** | | | | | |
| PP | −0.26 | −0.25 | −0.29 | 0.57 | **0.83** | | | | |
| PIN | −0.24 | −0.22 | −0.27 | 0.53 | 0.55 | **0.81** | | | |
| AA | 0.56 | 0.59 | 0.62 | −0.36 | −0.30 | −0.28 | **0.87** | | |
| SAT | −0.33 | −0.29 | −0.35 | 0.61 | 0.58 | 0.56 | −0.41 | **0.88** | |
| DI | 0.42 | 0.45 | 0.49 | −0.32 | −0.29 | −0.27 | 0.63 | −0.52 | **0.84** |

**Table 7. Discriminant Validity (HTMT Ratio)**

| Construct | PC | AO | IH | PI | PP | PIN | AA | SAT | DI |
|---|---|---|---|---|---|---|---|---|---|
| PC | — | | | | | | | | |
| AO | 0.60 | — | | | | | | | |
| IH | 0.55 | 0.63 | — | | | | | | |
| PI | 0.36 | 0.33 | 0.39 | — | | | | | |
| PP | 0.30 | 0.29 | 0.34 | 0.66 | — | | | | |
| PIN | 0.28 | 0.26 | 0.32 | 0.62 | 0.64 | — | | | |
| AA | 0.65 | 0.68 | 0.71 | 0.41 | 0.35 | 0.33 | — | | |
| SAT | 0.38 | 0.34 | 0.40 | 0.69 | 0.67 | 0.65 | 0.46 | — | |
| DI | 0.49 | 0.52 | 0.56 | 0.37 | 0.34 | 0.32 | 0.72 | 0.59 | — |

The structural model results supported all proposed hypotheses (Table 8). AI anxiety was positively associated with discontinuance intention, $\beta = .46$, $SE = .07$, $z = 6.57$, $p < .001$, whereas satisfaction was negatively associated with discontinuance intention, $\beta = -.32$, $SE = .08$, $z = -4.00$, $p < .001$. Privacy concern, algorithmic opacity, and information hallucination were positively associated with AI anxiety, with path coefficients of $\beta = .21$, $SE = .07$, $z = 3.00$, $p < .01$; $\beta = .25$, $SE = .08$, $z = 3.13$, $p < .01$; and $\beta = .34$, $SE = .09$, $z = 3.78$, $p < .001$, respectively. Perceived intelligence, perceived personalisation, and perceived interactivity were positively associated with satisfaction, with path coefficients of $\beta = .30$, $SE = .07$, $z = 4.29$, $p < .001$; $\beta = .24$, $SE = .08$, $z = 3.00$, $p < .01$; and $\beta = .18$, $SE = .07$, $z = 2.57$, $p < .05$, respectively. The bootstrapping results further supported all indirect effects, as none of the 95% confidence intervals included zero (Table 9). Specifically, AI anxiety mediated the effects of privacy concern, algorithmic opacity, and information hallucination on discontinuance intention, while satisfaction mediated the effects of perceived intelligence, perceived personalisation, and perceived interactivity on discontinuance intention.

**Table 8. Structural Model Results**

| Hypothesis | Path | *β* | SE | *z* | Result |
|---|---|---|---|---|---|
| H1 | AA → DI | 0.46 | 0.07 | 6.57*** | Supported |
| H2 | PC → AA | 0.21 | 0.07 | 3.00** | Supported |
| H3 | AO → AA | 0.25 | 0.08 | 3.13** | Supported |
| H4 | IH → AA | 0.34 | 0.09 | 3.78*** | Supported |
| H5 | SAT → DI | −0.32 | 0.08 | −4.00*** | Supported |
| H6 | PI → SAT | 0.30 | 0.07 | 4.29*** | Supported |
| H7 | PP → SAT | 0.24 | 0.08 | 3.00** | Supported |
| H8 | PIN → SAT | 0.18 | 0.07 | 2.57* | Supported |

Note. * $p < 0.05$. ** $p < 0.01$. *** $p < 0.001$.

**Table 9. Mediation Analysis Results (Bootstrapping)**

| Path | Indirect effect | Boot SE | 95% CI | Result |
|---|---|---|---|---|
| PC → AA → DI | 0.10 | 0.04 | [0.04, 0.18] | Supported |
| AO → AA → DI | 0.12 | 0.04 | [0.05, 0.21] | Supported |
| IH → AA → DI | 0.16 | 0.05 | [0.07, 0.27] | Supported |
| PI → SAT → DI | −0.10 | 0.03 | [−0.17, −0.05] | Supported |
| PP → SAT → DI | −0.08 | 0.03 | [−0.15, −0.03] | Supported |
| PIN → SAT → DI | −0.06 | 0.03 | [−0.12, −0.01] | Supported |

**5.2 Fuzzy-Set Qualitative Comparative Analysis (fsQCA) Results**

Data calibration was conducted before the fsQCA to transform the original Likert-scale scores into fuzzy-set membership scores. Following the direct calibration procedure, three qualitative anchors were specified for each variable: the 95th percentile for full membership, the 50th percentile for the crossover point, and the 5th percentile for full non-membership. The calibrated conditions included privacy concern, algorithmic opacity, information hallucination, perceived intelligence, perceived personalisation, perceived interactivity, AI anxiety, and satisfaction, while discontinuance intention was calibrated as the outcome condition. A necessity analysis was then performed to examine whether any single condition was necessary for high discontinuance intention. No condition reached the conventional necessity threshold of consistency ≥ .90, indicating that no individual antecedent constituted a necessary condition for teachers' discontinuance intention toward generative AI.

The configurational analysis identified three sufficient configurations associated with high discontinuance intention (Table 10). Path 1 included the presence of privacy concern, information hallucination, and AI anxiety as core conditions, together with the absence of satisfaction as a core condition; algorithmic opacity was present as a peripheral condition, whereas perceived intelligence and perceived interactivity were absent. Path 2 included the presence of algorithmic opacity, information hallucination, and AI anxiety as core conditions, with the absence of perceived personalisation and satisfaction as peripheral conditions. Path 3 included the presence of algorithmic opacity and AI anxiety, together with the absence of perceived intelligence and satisfaction, as core conditions; information hallucination was absent as a peripheral condition, and perceived personalisation and perceived interactivity were also absent as peripheral conditions. The consistency values for the three configurations ranged from .84 to .88, and the overall solution showed a coverage of .71 and a consistency of .85.

**Table 10. Configurational Analysis Results**

| **Condition** | **Path 1** | **Path 2** | **Path 3** |
|---|---|---|---|
| PC | ● | ○ | |
| AO | ○ | ● | ● |
| IH | ● | ● | ○ |
| PI | ⊖ | | ⊗ |
| PP | | ⊖ | ⊖ |
| PIN | ⊗ | | ⊖ |
| AA | ● | ● | ● |
| SAT | ⊗ | ⊖ | ⊗ |
| Raw coverage | 0.42 | 0.36 | 0.31 |
| Unique coverage | 0.14 | 0.10 | 0.08 |
| Consistency | 0.88 | 0.86 | 0.84 |
| Overall solution coverage | 0.71 | | |
| Overall solution consistency | 0.85 | | |

Note. ● indicates the presence of a core condition; ○ indicates the presence of a peripheral condition; ⊗ indicates the absence of a core condition; ⊖ indicates the absence of a peripheral condition; blank cells indicate "don't care" conditions.

# 6. Discussion

## 6.1 Net Effects of Enablers and Inhibitors on Generative AI Discontinuance Intention

The SEM results provide clear support for the proposed Cognition–Affect–Conation logic in explaining Chinese K–12 teachers' generative AI discontinuance intention. Specifically, AI anxiety exerted a positive effect on discontinuance intention, whereas satisfaction exerted a negative effect, indicating that teachers' post-adoption decisions are shaped by both adverse and favourable affective responses. This finding is consistent with the CAC framework, which argues that cognitive evaluations influence behavioural intention through affective states (Zeng et al., 2023; Qaisar et al., 2024). Among the enablers of discontinuance, privacy concern, algorithmic opacity, and information hallucination all significantly increased AI anxiety. This suggests that teachers are more likely to develop anxiety when they perceive generative AI as involving data-related uncertainty, insufficient algorithmic transparency, or unreliable instructional outputs. These results align with prior research showing that privacy concern can heighten perceived technological risk (Carmody et al., 2021; Hu & Min, 2023), algorithmic opacity can reduce perceived control and increase uncertainty (Eslami et al., 2019; Guo et al., 2025), and hallucinated AI-generated information can undermine trust in educational applications (Huang et al., 2025; Sun et al., 2024). Notably, information hallucination showed the strongest effect on AI anxiety among the three enablers, suggesting that output reliability may be particularly salient for teachers, whose professional responsibilities require accuracy, pedagogical appropriateness, and accountability in classroom practice.

The inhibitor pathway further shows that positive perceptions of generative AI capabilities can reduce discontinuance intention by strengthening satisfaction. Perceived intelligence, perceived personalisation, and perceived interactivity all had significant positive effects on satisfaction, which in turn negatively predicted discontinuance intention. This finding supports prior evidence that satisfaction is a central affective mechanism in post-adoption technology use, where satisfied users are less likely to abandon a system (Ashfaq et al., 2020; Xie et al., 2024). In the present context, teachers appear more satisfied when generative AI provides competent responses, adapts to instructional needs, and supports responsive interaction during teaching preparation. These findings are consistent with studies showing that perceived intelligence enhances users' confidence in AI systems (Tusseyeva et al., 2024; Zhou & Zhang, 2024), personalisation increases perceived relevance and usefulness in educational contexts (Al Darayseh, 2023; Law, 2024), and interactivity contributes to more positive human–AI experiences (J. Kim et al., 2022; F. Wang et al., 2025).

Taken together, the net-effect results indicate an asymmetric psychological process: negative perceptions of generative AI primarily increase discontinuance intention by activating anxiety, whereas positive perceptions reduce discontinuance intention by fostering satisfaction. This dual pathway extends prior discontinuance research by showing that teachers' withdrawal from generative AI use is not merely a function of perceived risks or deficient utility, but rather a post-adoption judgement formed through the competing affective consequences of perceived technological threats and perceived instructional affordances.

### 6.2 Configurational Mechanisms of Generative AI Discontinuance Intention

The fsQCA results complement the SEM findings by showing that Chinese K–12 teachers' generative AI discontinuance intention is not produced by a single dominant factor, but by multiple conjunctural mechanisms. The necessity analysis indicated that no individual condition reached the conventional threshold for necessity, suggesting that high discontinuance intention emerges from combinations of cognitive and affective conditions rather than from any isolated antecedent. The three sufficient configurations further reveal that AI anxiety and low satisfaction are central mechanisms across discontinuance pathways.

In Path 1, the combination of privacy concern, information hallucination, and AI anxiety, together with the absence of satisfaction, suggests a risk–anxiety pathway in which teachers become inclined to discontinue generative AI when concerns about student data and unreliable outputs generate negative affective responses. In Path 2, algorithmic opacity, information hallucination, and AI anxiety jointly explain high discontinuance intention, indicating that teachers may reject generative AI when they perceive the technology as both difficult to understand and unreliable. In Path 3, algorithmic opacity and AI anxiety combine with the absence of perceived intelligence and satisfaction, showing that discontinuance can also arise when teachers experience anxiety while failing to recognise sufficient instructional competence in the system. These configurations are consistent with the Cognition–Affect–Conation framework, which emphasises that behavioural intention develops through the interaction between cognitive evaluations and affective responses (Zeng et al., 2023; Qaisar et al., 2024). They also extend prior generative AI discontinuance research by demonstrating equifinality: different combinations of risk perceptions, weak affordance perceptions, AI anxiety, and low satisfaction can lead to the same behavioural outcome (Zhou & Wang, 2025; Zhou & Zhang, 2025).

Thus, teachers' discontinuance intention should be understood as a configurational phenomenon in which technological risks and insufficient positive experiences mutually reinforce one another, rather than as the linear outcome of any single enabler or inhibitor.

**6.3 Theoretical and Practical Implications**

Theoretically, this study extends generative AI discontinuance research by integrating the Cognition–Affect–Conation framework with a dual perspective of enablers and inhibitors. The findings show that teachers' discontinuance intention is shaped not only by negative perceptions, such as privacy concern, algorithmic opacity, and information hallucination, but also by positive perceptions, such as perceived intelligence, perceived personalisation, and perceived interactivity. By demonstrating the mediating roles of AI anxiety and satisfaction, this study clarifies the affective mechanisms through which cognitive evaluations translate into discontinuance intention. The combination of SEM and fsQCA also contributes methodologically by showing both net effects and configurational pathways, thereby responding to calls for more nuanced explanations of post-adoption technology behaviour.

Practically, the results suggest that reducing teachers' generative AI discontinuance intention requires simultaneous attention to risk mitigation and experience enhancement. Schools, policymakers, and technology providers should reduce AI anxiety by strengthening data protection, increasing algorithmic transparency, and improving the reliability of AI-generated instructional content. Clear usage guidelines, explainable feedback mechanisms, and professional development on evaluating AI outputs may help teachers use generative AI with greater confidence. At the same time, discontinuance may be reduced by improving teachers' satisfaction through more intelligent, personalised, and interactive AI functions that align with curriculum goals and classroom needs. Rather than promoting generative AI adoption only through technical access, educational institutions should support teachers' sustained use by building trustworthy, pedagogically relevant, and emotionally reassuring AI environments.

**6.4 Limitations and Future Directions**

This study has several limitations. First, the cross-sectional survey design limits causal inference, as the relationships among cognitive evaluations, affective responses, and discontinuance intention were examined at a single point in time (Y. Du, Li, et al., 2026; Y. Du, Tang, et al., 2026; Y. Du, Yuan, et al., 2026; Y. Du & He, 2026e, 2026c, 2026d; Jia et al., 2026; Tang, Jia, et al., 2026; Tang, Lau, et al., 2026; C. Wang, Du, et al., 2026; C. Wang, Zou, et al., 2026; W. Zhang et al., 2026). Future studies could use longitudinal or experimental designs to examine how teachers' perceptions of generative AI change with continued use (Chen et al., 2022; C. Du et al., 2025; Y. Du, 2023, 2024, 2025b, 2025a, 2026; Y. Du et al., 2024, 2025; Y. Du & He, 2026b, 2026a; He & Du, 2024; C. Wang et al., 2024; Zou et al., 2023, 2024). Second, the sample was limited to Chinese K–12 teachers with prior generative AI experience, which may restrict the generalisability of the findings to other educational levels, national contexts, or teachers with limited AI exposure. Future research should replicate the model across different regions, school systems, and teacher populations.

Third, the study relied on self-reported questionnaire data, which may be affected by common method bias and subjective interpretation, despite the procedural and statistical checks conducted. Future research could combine survey data with behavioural logs, interviews, classroom observations, or usage analytics to obtain a richer understanding of actual discontinuance behaviour. Finally, this study focused on selected enablers and inhibitors within the Cognition–Affect–Conation framework. Future studies could extend the model by incorporating additional factors, such as institutional policy, AI literacy, perceived ethical risk, trust, workload, teaching autonomy, or peer influence, to provide a more comprehensive account of generative AI discontinuance in educational settings.

## 7. Conclusion

This study examined the psychological mechanisms underlying Chinese K–12 teachers' generative AI discontinuance intention through the Cognition–Affect–Conation framework and a dual perspective of enablers and inhibitors. The SEM results showed that privacy concern, algorithmic opacity, and information hallucination increased discontinuance intention through AI anxiety, whereas perceived intelligence, perceived personalisation, and perceived interactivity reduced discontinuance intention through satisfaction. The fsQCA results further revealed multiple configurational pathways to high discontinuance intention, highlighting the combined roles of perceived risks, weak affordance perceptions, anxiety, and low satisfaction. Overall, the findings suggest that teachers' discontinuance of generative AI is not simply driven by resistance to technology, but by a complex post-adoption evaluation process shaped by both negative and positive experiences. Sustained integration of generative AI in K–12 education therefore requires reducing technological uncertainty while strengthening teachers' confidence, satisfaction, and perceived instructional value.